\documentclass[%
reprint,
 amsmath,amssymb,
 aps,
]{revtex4-2}

\usepackage{graphicx}
\usepackage{dcolumn}
\usepackage{bm}
\usepackage{color}

\newtheorem{theorem}{Theorem}[section]

\newtheorem{corollary}[theorem]{Corollary}

\newtheorem{example}[theorem]{Example}

\def\IR{{\bf R}}
\def\IC{{\bf C}}
\def\IR{{\rm I\!R}}

\def\la{{\langle}}
\def\ra{{\rangle}}

\def\diag{{\rm diag}\,}
\def\tr{{\rm tr}\,}

\def\qed{\hfill\vbox{\hrule width 6 pt\hbox{\vrule height 6 pt width 6
pt}}\smallskip}

\def\bM{{\bf M}}

\def\IR{{\mathbb R}}
\def\IC{{\mathbb C}}

\begin{document}


\title{
Unified Derivation of Uncertainty Relations and Their Saturation Conditions}

\author{Chi-Kwong Li}
 \email{ckli@math.wm.edu}
\affiliation{%
Department of Mathematics, College of William \& Mary,
Williamsburg, VA 23187, USA
}%

\author{Mikio Nakahara}
\email{mikio.nakahara@meetiqm.com}
\affiliation{IQM Quantum Computers, Keilaranta 19
02150 Espoo, Finland
}%
\affiliation{
Research Institute for Science and Technology,
Kindai University
Higashi-Osaka, 577-8502 JAPAN
}%

\date{\today}

\begin{abstract}
We analyze uncertainty relations due to Kennard, Robertson, 
Schr\"odinger, Maccone and Pati in a unified way from matrix theory point of view.
Short proofs are given to these uncertainty relations 
and characterizations of the saturation conditions are given.
\end{abstract}

\maketitle







Keywords.  Quantum states, Uncertainty relations, Observables

\section{Introduction}

Uncertainty relation is one of the most distinguished properties of quantum mechanics
compared to its classical counterpart. It is impossible to measure two mutually
non-commutative observables simultaneously with infinite precision.
This uncertainty was formulated in many ways. Heisenberg first proposed a Gedanken
experiment, in which both position and momentum are measured simultaneously
and discussed the effect of one measurement on the other.
Later, Robertson and Kennard formulated the uncertainty principle in terms of
standard deviations of measurement outcomes with respect to the state under 
consideration. Subsequently, Schr\"odinger proposed a stronger inequality involving
expectation value of anti-commutation relations.
These uncertainty relations describe inequality between the product of
the standard deviations associated with two observables and expectation values of
the commutator and anti-commutator of these observables. Recently
Maccone and Pati proposed a new type of uncertainty relations that involve
a {\it sum} of two standard deviations squared. One may see 
\cite{Kennard,MP,Robertson,S} for a
general background.  It is the purpose of this Letter to
derive these uncertainty relations in a unified manner and characterize
the saturation conditions, i.e., the conditions when the lower bounds
are attained.

\smallskip
In the following we consider a quantum system whose Hilbert space is $\IC^n$ and 
consider a pure state $|\psi\rangle \in \IC^n$ and
a mixed state $\rho \in \bM_n$, where $\bM_n$ denotes the set of 
$n\times n$ complex matrices.  Let $a$ be an observable and $A \in \bM_n$ be the corresponding Hermitian matrix. The expectation value $\alpha$ and the standard deviation $\Delta(a)$
of $a$ is defined as follows.
\smallskip\noindent
(1)  For a pure state $|\psi\ra \la \psi|$, 
$\alpha = \la \psi|A|\psi \ra$
and 
$$
\Delta(A) = \sqrt{\langle \psi|
(A-\alpha I)^2|\psi\rangle} 
=\sqrt{\langle \psi |
A^2|\psi\rangle - \alpha^2}.
$$
(2) For a mixed state $\rho$,
$\alpha = \tr (\rho A)$ 
and 
$$\Delta(A) =  \sqrt{ \tr  (\rho(A-\alpha I)^2)} = 
\sqrt{\tr (\rho A^2) -\alpha^2}.$$  
We use the same
symbols $\alpha$ and $\Delta(A)$ both for a pure state and a mixed state;
which case is under consideration should be obvious from the context.

\section{Uncertainty principles by Robertson and Schr\"odinger}

It is convenient to introduce $\tilde  A = A - \alpha I$ in the following discussion.
Note that 
$\la \psi|\tilde  A| \psi \ra= \tr (\tilde  A \rho) =0,$
$$
\Delta(A) = \Delta (\tilde  A) = \sqrt{\la \psi|\tilde  A^2|\psi \ra} \ \hbox{ for a pure state},$$
and
$$
\Delta(A) = 
\Delta (\tilde  A)= \sqrt{\tr (\tilde  A^2 \rho)} \quad \hbox{ for a mixed state}.
$$
For another Hermitian matrix $B\in \bM_n$, 
let $\beta = \la \psi|B|\psi\ra$, $\tilde B = B- \beta I$,
and   $\Delta(B)$ be the standard deviation.

We begin with the uncertainty principle proposed by Robertson \cite{Robertson}.
 
\begin{theorem} \label{2.1}
Let $A, B\in \bM_n$ be Hermitian, $|\psi\rangle \in \IC^n$ 
be a unit vector. 
Then
\begin{equation}\label{uncertainty}
\Delta(A)\Delta(B) \ge |\langle \psi |
[A, B]|\psi\rangle|/2,
\end{equation}
where $[A, B]=AB-BA$.
The equality holds in {\rm (\ref{uncertainty})} if and only if 
there is $\vartheta \in [0, 2\pi)$ such that
$$(\cos\vartheta A + i \sin \vartheta B)|\psi\ra 
= (\alpha \cos\vartheta  + i \beta\sin\vartheta)|\psi\ra.$$
\end{theorem}

\it Proof.  \rm
Since
$\Delta(A)\Delta(B) = 
\sqrt{\langle \psi |
\tilde  A^2 |\psi\rangle}\sqrt{ \langle \psi |
\tilde  B^2 |\psi\rangle}$ and
$\langle \psi |
[A, B]|\psi\rangle = \langle \psi |
[\tilde  A, \tilde  B]|\psi\rangle$, 
it suffices to show that 
$$4\langle \psi |
\tilde  A^2|\psi\rangle \langle \psi |
\tilde  B^2|\psi\rangle  
\ge | \langle \psi |
[\tilde  A, \tilde  B]|\psi\rangle|^2.$$
The matrices
$$C_1  = \begin{pmatrix} \langle \psi |
\tilde  A^2|\psi\rangle  &\langle \psi |
\tilde  A\tilde  B|\psi\rangle \cr 
\langle \psi |
\tilde  B\tilde  A|\psi\rangle & \langle \psi |
\tilde  B^2|\psi\rangle \cr\end{pmatrix}$$ 
and
$$C_2 = \begin{pmatrix} \langle \psi |
\tilde  A^2|\psi\rangle  & -\langle \psi |
\tilde  B\tilde  A|\psi\rangle \cr 
-\langle \psi |
\tilde  A\tilde  B|\psi\rangle & \langle \psi |
\tilde  B^2|\psi\rangle \cr\end{pmatrix} $$
are Hermitian, and $\det(\lambda I - C_1) = \det(\lambda I-C_2)=0$
has solutions
\begin{eqnarray*} &&
\frac{1}{2} \left[\|\tilde A|\psi\ra \|^2+\|\tilde B|\psi\ra \|^2 \right.\\
&& \left. \pm \sqrt{(\|\tilde A|\psi\ra \|^2+
\|\tilde B|\psi\ra \|^2)^2 - 4|\la \psi |\tilde A \tilde B|\psi\ra|^2}\right],
\end{eqnarray*}
which are non-negative as
$$(\|\tilde A|\psi\ra 
\|^2+\|\tilde B|\psi\ra \|^2)^2 - 4|\la \psi |\tilde A \tilde B|\psi\ra|^2 \ge 0$$
by the Cauchy-Schwartz inequality.  
Thus, $C_1, C_2$, and $C = C_1 + C_2$ are positive semi-definite
and 
$$4\langle \psi |
\tilde  A^2|\psi \rangle \langle \psi |
\tilde  B^2|\psi \rangle 
 - |\langle \psi |
[\tilde  A, \tilde  B]|\psi \rangle|^2 = \det(C) \ge 0.$$
The equality $\det(C) = 0$ holds if and only if  $C$ is singular, equivalently,  
the positive semi-definite matrices $C_1$ and $C_2$ are singular and  
share a common null vector, which will be a null vector of $C$. 
Since $C_1$ and $C_2$ have the same trace, we see that $C_1$ and $C_2$ 
have a common null vector if and only if there is a unit vector
$|u\ra \in \IC^n$ such that
\begin{equation}\label{eq:2.2}
C_1 = C_2 = (\mathrm{tr}~C_1) |u\ra \la u|.
 \end{equation}
If (\ref{eq:2.2}) holds, then 
$$0 = \det(C_1) =  \langle \psi |
\tilde  A^2|\psi \rangle \langle \psi |
\tilde  B^2|\psi \rangle 
- |\langle \psi |
 \tilde  A\tilde  B |\psi \rangle|^2.$$
By the Cauchy-Schwartz inequality,
$\tilde  A|\psi \rangle$ and $\tilde  B|\psi \rangle$ are linearly dependent.
Since the $(1,2)$ entries of $C_1$ and $C_2$ are the same, i.e., 
$\langle \psi |\tilde  A\tilde  B|\psi \rangle = -\langle \psi |
\tilde  B\tilde  A|\psi \rangle$, 
it follows that $\langle \psi | \tilde  A\tilde  B|\psi \rangle \in i\IR.$
Hence,
$\tilde  A|\psi \rangle$ and $i\tilde  B|\psi \rangle$ are linearly 
dependent over $\IR$, and  there is $\vartheta \in [0, 2\pi)$ such that
$\cos\vartheta \tilde  A |\psi \rangle + i \sin \vartheta \tilde  B |\psi \rangle$ 
is the zero vector. Equivalently,
$$\cos\vartheta (A-\alpha I)  |\psi\rangle  + i\sin\vartheta( B-\beta I) |\psi\rangle = 0.$$

\smallskip
Conversely, if $\cos\vartheta \tilde  A |\psi \rangle + 
i \sin \vartheta \tilde  B |\psi \rangle$ 
is the zero vector, then for $|v\ra = (\cos \vartheta, i \sin \vartheta)^t$,
we have $\la v|C_1| v\ra = \la v|C_2|v\ra = 0$ so that $\la v|C|v\ra = 0$.
Thus, $|v\ra$ is a null vector of the positive semi-definite matrix $C$
and $\det(C) = 0$. Thus the inequality 
{\rm (\ref{uncertainty}) becomes an equality. \qed

It is often said that the uncertainty problem for a finite dimensional system is trivial since an eigenvector of one of the operators $A$ and $B$ saturates the inequality so that the both sides vanish. This is the case when $\vartheta=0$ ($\hat A |\psi\rangle=0$) or $\vartheta=\pi/2$ ($\hat B|\psi \rangle=0$), see Section IV. We have shown here that there are states that saturate the inequality non-trivially so that the both sides are a non-negative number.

\begin{example} \label{2.2} Let $A =\sigma_x$ 
and $B  = \sigma_y$ be the Pauli matrices and $|\psi \ra
= (\cos(\theta/2), e^{i \phi} \sin(\theta/2))^t$.
Then $\alpha = \sin \theta \cos \phi$ and $\beta = \sin \theta
\sin \phi$. Let
$$
\cos \vartheta \hat{A} |\psi \ra + i \sin \vartheta \hat{B}|\psi \ra
=0
$$
and find non-trivial solutions of this equation. 
The first component vanishes if and only if $\sin \vartheta-
\cos \vartheta \cos \theta =0$ and $\cos \vartheta
- \sin \vartheta \cos \theta =0$ simultaneously, namely
$\theta =0, \vartheta=\pi/4$ or $\theta = \pi, \vartheta=-\pi/4$. The second component vanishes under these conditions.
We find the solutions $|0 \ra 
= (1,0)^t$ and $|1 \ra = (0,1)^t$. 
Both sides of Eq.~{\rm(\ref{uncertainty})} are equal to 1 for these solutions.
\end{example}

\smallskip
More generally, a quantum system in a mixed state 
is represented by a density matrix $\rho\in \bM_n$.
Then the expectation value and the standard deviation of an
observable $a$ associated with the Hermitian matrix $A\in \bM_n$ 
are $\alpha = \tr (A \rho)$ and 
$$
\Delta(A) = \Delta(\tilde A) = \sqrt{\tr(\tilde  A^2 \rho)} = 
\sqrt{\tr (\rho^{1/2}\tilde  A^2\rho^{1/2})}. $$
Let $\la X,Y\ra = 
\tr (X^{\dagger} Y)$ be the Frobenius inner product for $X, Y \in \bM_n$.
Then   
\begin{equation} \label{m-dev}
\Delta(A) = \Delta(\tilde A) = \sqrt{\la \tilde  A \rho^{1/2}, \tilde  A \rho^{1/2} \ra}.
\end{equation}
If $\rho = |\psi \rangle\langle \psi|$, then 
the expectation value and 
$\Delta(A)$ reduce to the previous definition.
We have the following.

\smallskip
\begin{theorem} \label{2.3}
Let $\rho \in \bM_n$ be a density matrix,
and let $A, B \in \bM_n$ be Hermitian matrices.
Then 
$$\Delta(A) \Delta(B) \ge |\tr([A, B] \rho)|/2.$$
The equality holds if and only if there is $\vartheta \in [0, 2\pi)$ such that
\begin{equation}
\label{eq2}
\cos \vartheta (A-\alpha I) \rho^{r} + i \sin \vartheta (B-\beta I) \rho^{r} = 0
\end{equation}
for any positive number $r$.
\end{theorem}

\it Proof. \rm 
We again exclude the trivial cases where $\rho$ is a pure state and $\vartheta=0, A \rho = \alpha \rho$ and $\vartheta=\pi/2, B \rho=\beta \rho$. 

Note that 
$\Delta(A) \Delta(B) = \Delta(\tilde A) \Delta(\tilde B)$ 
and $[A, B]= [\tilde A, \tilde B]$. 
We adapt the proof of Theorem \ref{uncertainty} and consider
$$C_1  = \begin{pmatrix} \la \tilde  A \rho^{1/2}, \tilde  A \rho^{1/2}\ra  
&\la\tilde  A \rho^{1/2}, \tilde  B \rho^{1/2}\ra \cr 
\la\tilde  B \rho^{1/2}, \tilde  A \rho^{1/2}\ra & 
\la \tilde  B \rho^{1/2}, \tilde  B \rho^{1/2}\ra\cr
\end{pmatrix}$$
and 
$$
C_2  = \begin{pmatrix} \la\tilde  A \rho^{1/2}, \tilde  A \rho^{1/2}\ra 
& - \la\tilde  B \rho^{1/2}, \tilde  A \rho^{1/2}\ra\cr 
-\la\tilde  A \rho^{1/2}, \tilde  B \rho^{1/2}\ra & 
\la\tilde  B \rho^{1/2}, \tilde  B \rho^{1/2}\ra\cr
\end{pmatrix}.$$
Then $C_1$ and $C_2$ are positive semi-definite matrices
and so is $C_1+C_2$.   Thus, $\det(C_1+C_2) \ge 0$, and 
\begin{eqnarray*}
&&\Delta(A) \Delta(B) = \Delta(\tilde A) \Delta(\tilde B) \\
&\ge& |\tr([\tilde A, \tilde B] \rho)|/2 = 
|\tr([A, B] \rho)|/2.\end{eqnarray*}

\smallskip
The equality $\det(C_1+C_2) = 0$ holds if and only if 
$C_1 = C_2 = (\tr C_1)|u\ra\la u|$
for a unit vector $|u\ra \in \IC^2$. Equivalently,
$$\la\tilde  A \rho^{1/2}, \tilde  A \rho^{1/2}\ra 
\la\tilde  B \rho^{1/2}, \tilde  B \rho^{1/2}\ra = |\la\tilde  A \rho^{1/2}, \tilde  B \rho^{1/2}\ra|^2$$
and 
$$ \la\tilde  A \rho^{1/2}, \tilde  B \rho^{1/2}\ra = 
-\la\tilde  B \rho^{1/2}, \tilde  A \rho^{1/2}\ra.$$
These equalities hold if and only if $\tilde A \rho^{1/2}$ and$\tilde B \rho^{1/2}$
are linearly dependent in $\bM_n$ and there are $\vartheta \in [0, 2\pi)$
such that
$$0  
= \cos \vartheta \tilde  A \rho^{1/2} + i \sin \vartheta \tilde  B \rho^{1/2}.$$
Equivalently,
\begin{equation}\label{eq3}
0= \cos \vartheta (A-\alpha I) \rho^{1/2} + i \sin \vartheta (B-\beta I) \rho^{1/2}.
\end{equation}

\smallskip
We will show that (\ref{eq3}) holds if and only if  
condition (\ref{eq2}) holds. Suppose (\ref{eq3}) holds and $U$ is unitary such that 
$U^\dag
\rho^{1/2}U = D \oplus 0_{n-k}$, where
$D$ is a diagonal matrix with positive diagonal 
entries and $0_{n-k}\in \bM_{n-k}$ is the zero matrix. 
If $U^\dag \tilde  A U = (A_{ij})$ and $U^\dag \tilde  B U = (B_{ij})$
with $1 \le i,j \le 2$ such that $A_{11}, B_{11} \in \bM_k$.
So, 
$$ \begin{pmatrix} (\cos\vartheta A_{11} + i\sin\vartheta B_{11}) D & 0 \cr
(\cos\vartheta A_{21}+i\sin\vartheta B_{21}) D& 0_{n-k}\cr\end{pmatrix} = 0.$$
Since $D$ is invertible, we see that $\cos\vartheta A_{11}+i\sin \vartheta B_{11}$ and 
$\cos\vartheta A_{21}+i\sin \vartheta B_{21}$ are zero matrices.
Since $A_{11}, B_{11}$ are Hermitian, we see that
$A_{11} = B_{11} = 0$. The $(2,1)$ block satisfies
$\cos\vartheta A_{21} + i \sin \vartheta B_{21} = 0_{n-k,k}$.
By multiplying $\rho^{r-1/2}$ to (\ref{eq3}) from the right we find
$$
\cos \vartheta \tilde  A \rho^r + i \sin \vartheta \tilde  B \rho^r=0. 
$$
Note that $\rho^r$ cannot vanish since $\rho^r = U(D^{2r} \oplus 0_{n-k})U^{\dagger} $
for any positive $r$.

Conversely, if (\ref{eq2}) is satisfied, then (\ref{eq3}) is obtained by multiplying 
$U (D^{1-2r} \oplus 0_{n-k})U^{\dagger}$ from the right. 
Here, $D^{1-2r}$ is well defined as $D$ is positive definite.
\qed

\smallskip
If we let $r  =1$ in  (\ref{eq2}), then 
the  condition reduces to 
\begin{equation}
\label{eq4}
\cos \vartheta (A-\alpha I)\rho + i\sin \vartheta (B-\beta I)\rho  = 0.
\end{equation}

If $\rho = |\psi\ra \la \psi|$ is a pure state, then we can multiply the
$|\psi\ra$ to (\ref{eq4}) on the right to obtain  
the result in Theorem \ref{2.1}.
In fact, if $n=2$, the equality is satisfied only by pure states, independently of $A$ and $B$, as the following corollary proves.

\begin{corollary}
For a qubit state $\rho \in \bM_2$, and  non-scalar Hermitian matrices $A, B \in \bM_2$,
{\rm (\ref{eq4})} can only happen  for a pure state $\rho$.
\end{corollary}

\it Proof. \rm  
If (\ref{eq4}) holds and if $\rho \in \bM_2$ has rank 2, then 
$\rho^{r}$ is invertible. Hence
$\cos\vartheta \tilde A + i\sin\vartheta \tilde B=0$.
Since $\tilde A$ and $\tilde B$ are Hermitian,
$\tilde A = \tilde B = 0$, i.e., $A = \alpha I, B = \beta I$, 
which is a contradiction.\qed

For higher dimensions, we may have the following.

\begin{example} Let
$A = \begin{pmatrix} 0_2& I_2\cr I_2 & 0_2\cr\end{pmatrix},
B = \begin{pmatrix} 0_2& -iI_2\cr iI_2 & 0_2\cr\end{pmatrix} \in M_4$
and $\rho = I_2/2 \oplus 0_2$.
Then $(\tilde  A, \tilde  B) = (A,B)$ and
$$\tr (A^2 \rho) \tr(B^2\rho) = |\tr([A,B]\rho)|^2 = 1.$$ 
\end{example}


Next, we consider the uncertainty principle proposed by
Schr\"{o}dinger \cite{S}, which is an improvement of the one by Robertson.
We consider the general form on mixed states, and let 
$\Delta(A)$ be defined as in (\ref{m-dev}) in the following.

\begin{theorem} 
Let $\rho \in \bM_n$ be a density matrix,  $A, B \in \bM_n$ be Hermitian matrices.
Then
\begin{equation}
\label{Schro}
\Delta(A)^2\Delta(B)^2 \ge \left|\frac{1}{2} \tr(\{A, B\}\rho)-\alpha\beta\right|^2 + 
\left|\frac{1}{2i} \tr([A,B]\rho)\right|^2.
\end{equation}
The inequality become equality if and only if for any $r >0$, the 
two matrices
$(A-\alpha I) \rho^r, (B-\beta I) \rho^r$
are linearly dependent,  i.e., there are $\vartheta, \varphi \in \IR$
satisfying 
$$
[\cos \vartheta (A-\alpha I) + 
e^{i\varphi} \sin \vartheta (B-\beta I)]\rho^r = 0.$$
\end{theorem}

\it Proof. \rm  Continue to let $(\tilde A, \tilde B) = (A-\alpha I, B - \beta I)$.
One readily checks that inequality (\ref{Schro}) is the same as
\begin{eqnarray*}
&&\la \tilde  A \rho^{1/2}, \tilde  A \rho^{1/2}\ra
\la \tilde  B \rho^{1/2}, \tilde  B \rho^{1/2}\ra \\
&\ge&  
\left|\frac{1}{2} \tr(\{\tilde  A, \tilde  B\}\rho)\right|^2 + 
\left|\frac{1}{2i} \tr([\tilde  A,\tilde  B]\rho)\right|^2.
\end{eqnarray*}
Since 
$$\left|\frac{1}{2} \tr(\{\tilde  A, \tilde  B\}\rho)\right|^2 + 
\left|\frac{1}{2i} \tr([\tilde  A,\tilde  B]\rho)\right|^2
=  |\la\tilde  A \rho^{1/2},\tilde  B \rho^{1/2}\ra|^2,$$
the above inequality follows from the fact that the matrix 
$C_1  = \begin{pmatrix} \la \tilde  A \rho^{1/2}, \tilde  A \rho^{1/2}\ra  
&\la\tilde  A \rho^{1/2}, \tilde  B \rho^{1/2}\ra \cr 
\la\tilde  B \rho^{1/2}, \tilde  A \rho^{1/2}\ra & 
\la \tilde  B \rho^{1/2}, \tilde  B \rho^{1/2}\ra\cr
\end{pmatrix}$ is positive semi-definite, and has non-negative determinant.

\smallskip
Moreover, inequality (\ref{Schro}) becomes equality if and only if $\det(C_1) = 0$, i.e.,
the two matrices
$\tilde  A\rho^{1/2}, \tilde  B\rho^{1/2}$ are linearly dependent. 
Using a similar argument as in the proof of Theorem \ref{2.3},
we see that 
the equality condition is equivalent that
the two matrices $\tilde   A \rho^r, \tilde  B \rho^r$ 
are linearly dependent for any $r >0$.
\qed

If we take $A= \sigma_x, B=\sigma_y$,  and $\rho = |\psi\ra \la \psi|$ 
as in Example \ref{2.2},
the solution we found for the Robertson inequality trivially satisfies 
the equality of the Schr\"odinger inequality since $\{A, B\} = 0$.

\section{Uncertainty principles by  Maccone and Pati}
\setcounter{equation}{0}

In this section, we study the uncertainty principles proposed 
by Maccone and Pati. First, we have the following refinement of 
the relation.

\begin{theorem} \label{3.1}
Let $|\psi \rangle \in \IC^n$ be a unit vector, $A, B \in \bM_n$ be  
Hermitian matrices, $(\alpha, \beta) = (\langle \psi |
A|\psi \rangle, \langle \psi |
B|\psi \rangle)$,
and $|\phi \rangle$ be a unit vector orthogonal to $|\psi \rangle$.  Then for any 
$\mu \in \IC$ with $|\mu| = 1$,
\begin{eqnarray*}
\Delta(A)^2 + \Delta(B)^2
&\ge& |\langle \psi | A |\phi \rangle|^2 + |\langle \psi | B |\phi \rangle|^2 \\
&\ge& \frac{1}{2}(|\langle \psi |A|\phi \rangle|+ 
|\langle \psi |B|\phi \rangle|)^2 \\
&\ge& \frac{1}{2}|\langle \psi |
(A + \mu B)|\phi \rangle|^2.
\end{eqnarray*}

\smallskip\noindent
$\bullet$
The first inequality becomes equality if and only if 
$$(\|(A-\alpha I) |\psi \rangle\|, \|(B-\beta I) |\psi \rangle\|)
= (|\langle \psi |A|\phi \rangle|, 
|\langle \psi |B|\phi \rangle|).$$
$\bullet$
The second inequality becomes equality if and only if 
$|\langle \psi | A|\phi \rangle| = |\langle \psi |B|\phi \rangle|$.

\smallskip\noindent
$\bullet$
The third inequality becomes equality if and only if there is $\vartheta \in [0, 2\pi)$ 
such that 
$$e^{i\vartheta} \langle \psi |(A+\mu B)|\phi \rangle 
=|\langle \psi |A|\phi \rangle| + |\langle \psi |B|\phi \rangle|.$$
Consequently, all the inequalities become equalities if and only if
there is $\vartheta \in [0, 2\pi)$ such that
\begin{eqnarray*}
&&\|(A-\alpha I)|\psi \rangle\| = \|(B-\beta I)|\psi \rangle\| = |\langle \phi |
A|\psi \rangle |
\\
&=& |\langle \phi |B|\psi \rangle| 
= e^{i\vartheta} \langle \psi | A|\phi \rangle = 
e^{i\vartheta} \langle \psi | \mu B|\phi \rangle;
\end{eqnarray*}
equivalently, $|\psi\ra$ is an eigenvector of $A-\mu^* B$ with eigenvalue 
$\alpha - \mu^* \beta$.
\end{theorem}

\it Proof. \rm
Let $U$ be unitary with $|\psi \rangle, |\phi \rangle $ as the first two columns, and 
$$U^\dag AU = \begin{pmatrix} \alpha & \langle  u| \cr | u \rangle & A_1\cr\end{pmatrix}
\quad \hbox{ and } \quad 
U^\dag BU = \begin{pmatrix} \beta & \langle v| \cr |v\rangle  & B_1\cr\end{pmatrix}.$$
Then  $c = \langle \psi |A|\phi \rangle, 
d= \langle \psi |B|\phi \rangle$ 
are the first entries of $\la u|$ and $\la v|$, respectively.
Observe that
$
\displaystyle
\langle \psi |(A+B)|\phi \rangle = c+d,$
$$
\langle \psi | \tilde  A^2 |\psi \rangle 
=\langle u | u\rangle = \| |u \rangle\|^2, \
\langle \psi | \tilde  B^2 |\psi \rangle
=\langle v|v \rangle= \| |v\rangle\|^2.$$
So, 
$$\frac{1}{2}|c+d|^2 \le 
\frac{1}{2}(|c| + |d|)^2  \le |c|^2+|d|^2 \le \|u\|^2 + \|v\|^2.$$ 
It is easy to check the condition for each  inequality becomes an equality, and
the last assertion concerning all the inequalities becomes equalities.
In particular, all the inequalities
become equalities if and only if $\la u| = \mu \la  v|$ has the 
form $(\gamma, 0, \dots, 0)$ with $\gamma = 
\la \psi|A|\phi\ra = \la \psi|\mu B|\phi\ra$,
i.e., $(A-\mu^* B)|\psi\ra = (\alpha-\mu^*\beta)|\psi\ra$.
\qed

\begin{example}
Let 

\smallskip\centerline{
$A=\sigma_x, B=\sigma_y$ and $|\psi \ra = (\cos (\theta/2),
e^{i \phi} \sin(\theta/2))^t$.}

\smallskip\noindent
Then 
$\alpha = \sin \theta
\cos \phi$ and $\beta = \sin \theta \sin \phi$. 
Let 
$$|\varphi \ra = (\sin(\theta/2), -e^{i \phi} \cos(\theta/2))^t$$
be a vector orthogonal to $|\psi \ra$. 

\smallskip\noindent
\begin{itemize}
\item[{\rm (1)}]
Then
\begin{eqnarray*}
\la \psi|\tilde A^2|\psi \ra&=& |\la \psi |\tilde A|\varphi\ra|^2=
\cos^2 \theta \cos^2 \phi + \sin^2 \phi,\\
\la \psi|\tilde B^2|\psi \ra&=& |\la \psi |\tilde B|\varphi\ra|^2
= \cos^2 \theta \sin^2 \phi+ \cos^2 \phi,
\end{eqnarray*}
showing the first equality is identically satisfied for any $|\psi \ra
\in \mathbb{C}^2$.

\item[{\rm (2)}]
The second equality is satisfied if and only if 
$$|\la \psi|\tilde A|\varphi \ra|= |\la \psi|\tilde B|\varphi \ra|,$$ 
which is written as
$\sin \theta \cos (2 \phi)=0$, i.e.,
$\theta=0, \pi$ or 
$\phi= \pm \pi/4,\pm  3\pi/4, \pm 5 \pi/4, \pm 7\pi/4.$

\item[{\rm (3)}] 
The third equality is satisfied with the above choices
of $(\theta, \phi)$ if we take
$$
\mu = \frac{\la \psi|A|\varphi \ra}{ \la \psi| B|\varphi \ra}
=\frac{-\sin \phi + i \cos \theta \cos \phi}
{\cos \phi + i \cos \theta \sin \phi}.
$$
In particular, $\mu= i$ for $\theta=0$ and $\mu=-i$ for $\theta=\pi$, but
$\mu$ does not take a simple form for other cases. 

\smallskip\noindent
From the above observations, we find the states satisfying all the
equalities in Theorem \ref{3.1} are

\begin{center}
$
(1,0)^t, (0,1)^t, (c, e^{\pm i \pi/4}s)^t,
(c, e^{\pm i 3\pi/4}s)^t,$

\smallskip
$(c, e^{\pm i5 \pi/4}s)^t,(c, e^{7\pm i \pi/4}s)^t,
$
\end{center}
where $c=\cos (\theta/2)$ and $s=\sin(\theta/2)$.
\end{itemize}
\end{example}

Next, we give short proofs and determine 
the saturation conditions for  two other uncertainty principles proposed in \cite{MP};
see Eqs.~(3) and (6) therein.

\begin{theorem} \label{3.3}
Suppose $|\psi\rangle\in \IC^n$ is a unit vector, $A, B\in \bM_n$ 
are Hermitian matrices, and 
$|\varphi\rangle\in \IC^n$ 
is a unit vector orthogonal to $|\psi\rangle$. 
Suppose $\mu  \in \{i, -i\}$ satisfies $\mu \la \psi|[A,B] |\psi \ra \ge 0$.
Then
\begin{equation}\label{MP3}
\Delta(A)^2 + \Delta(B)^2 
\ge \mu \langle \psi|[A,B]|\psi\rangle + |\langle \psi|A+\mu B|\varphi\rangle |^2,
\end{equation}
where the inequality becomes equality if and only if
\begin{equation}
\label{MP3equal}
\|((A-\alpha I)-\mu(B-\beta I))|\psi\ra\|   = |\la \psi |A+\mu B|\varphi\ra|.
\end{equation}

\smallskip
\noindent
Also, if $Q_\mu(A,B) = \left(\frac{A}{\Delta(A)} + \frac{\mu B}{\Delta(B)}\right)$, then
\begin{equation} \label{MP6}
\Delta(A)\Delta(B) 
\ge \frac{\mu}{2} \langle \psi|[A,B]|\psi\rangle /
\left(1-\frac{1}{2} \left| \langle \psi|Q_\mu(A,B)|\varphi\rangle \right|^2 \right),
\end{equation}
where the inequality becomes equality if and only if
\begin{equation}\label{MP6equal}
\left\|\left(\frac{A-\alpha I}{\sqrt{\Delta(A)}}-
\frac{\mu (B-\beta I)}{\sqrt{\Delta(B)}}\right)|\psi\ra\right\|   
= \left|\la \psi |Q_\mu(A,B)|\varphi\ra\right|.
\end{equation}
  \end{theorem}

\it Proof. \rm
Let $U$ be unitary with $|\psi \rangle, |\varphi \rangle $ as the first two columns, and 
let
$$U^\dag AU = \begin{pmatrix} \alpha & \langle  u| \cr | u \rangle & A_1\cr\end{pmatrix}
\quad \hbox{ and } \quad 
U^\dag BU = \begin{pmatrix} \beta & \langle v| \cr |v\rangle  & B_1\cr\end{pmatrix}.$$
Then $c = \langle \psi |A|\varphi \rangle$ and $d= \langle \psi |B|\varphi \rangle$ are
the first entries of $\la u|$ and $\la v|$, respectively, and
$$
\displaystyle
\langle \psi | \tilde  A^2 |\psi \rangle 
=\langle u | u\rangle = \| |u \rangle\|^2, \quad
\langle \psi | \tilde  B^2 |\psi \rangle
=\langle v|v \rangle= \| |v\rangle\|^2.$$ 
Since the first row of $U^*(\tilde  A + \mu \tilde  B)U$ equals
$\la u| + \mu \la v|$ with 
$c+\mu d$ as the first entry,
\begin{eqnarray*}
&&
\Delta(A)^2 + \Delta(B)^2 
- \mu \langle \psi|(AB-BA)|\psi\rangle\\
&=& 
\la u|u\ra + \la v | v\ra  -\mu (\la u|v\ra -  \la v|u \ra)\\
&=& \| |u-\mu v\ra \| = \| \la u|+\mu\la v|\|\\
&\ge& |c+\mu d|^2 = 
|\langle \psi|(A + \mu B)|\varphi\rangle |^2,
\end{eqnarray*}
which is (\ref{MP3}).
The equality holds if and only if 
$\la u - \mu  v| = e^{i\vartheta}(c+\mu d, 0, \dots, 0)$
for some $\vartheta \in [0, 2\pi)$. Equivalently, 
$|c+\mu d| =\| \la u|+\mu\la  v |\| = \| |u-\mu v \ra \| = 
\| ( \tilde A - \mu \tilde B)|\psi\ra\|$, i.e., (\ref{MP3equal}) holds.

\smallskip
To prove (\ref{MP6}), it suffices to show that
\begin{equation} \label{eq4.5}
1-\frac{1}{2} \left| \langle \psi|\left(\frac{A}{\Delta(A)} + 
\frac{\mu B}{\Delta(B)}\right)|\varphi\rangle \right|^2 
\ge \frac{\mu}{2} \frac{\langle \psi|(AB-BA)|\psi\rangle}{\Delta(A)\Delta(B)}.
\end{equation}
Using the previous notation, we have
\begin{equation}\label{eq5}
1 - \frac{1}{2}\left| \frac{c}{\| |u\ra\|} + \frac{\mu d}{\| |v\ra\|}\right|^2
\ge
\frac{\mu}{2}\frac{(\la u|v\ra - \la v|u\ra)}{\||u\ra\| \||v\ra\|}.
\end{equation}
Now, consider the unit vector $|\tilde u\ra = |u\ra/\|u\ra\|$ and 
$|\tilde v\ra = |v\ra/\|v\ra\|$. Then
$\frac{c}{\||u\ra\|} +\frac{\mu d}{\||v\ra\|}$ is  the  first entry of the 
row vector $\la \tilde u| + \mu \la \tilde v|
= \la \tilde u - \mu\tilde v|$. Hence
\begin{eqnarray*}
&& 2 - \frac{\mu(\la u|v\ra - \la v|u\ra)}{\||u\ra\| \||v\ra\|}\\
&=&  \| |\tilde u\ra \|^2 + \| |\tilde v\ra\|^2 
- \mu(\la \tilde u|\tilde v\ra - \la \tilde v|\tilde u\ra) \\ 
&=& \| |\tilde u - \mu \tilde v\ra\|^2  
\ge \left|\frac{c}{\||u\ra\|} +\frac{\mu d}{\||v\ra\|}\right|^2,
\end{eqnarray*}
which is equivalent to (\ref{eq5}). The equality holds if and only if
$\la \tilde u - \mu  \tilde v| 
= e^{i\vartheta}( \frac{c}{\||u\ra\|} +\frac{\mu d}{\||v\ra\|},
0, \dots, 0)$ for some $\vartheta \in [0, 2\pi)$. 
Equivalently, (\ref{MP6equal}) holds.
\qed

\smallskip
It  seems better to formulate the inequality as (\ref{eq4.5}) so that 
one does not need to worry about the situation when 
$1-\frac{1}{2} \left| \langle \psi|\left(\frac{A}{\Delta(A)} + 
\frac{\mu B}{\Delta(B)}\right)|\varphi\rangle \right|^2$ is zero
as long as $\Delta(A)$ and $\Delta(B)$ are nonzero.

\smallskip
By Theorem \ref{3.3}, it is easy to construct  an orthonormal set
$\{|\psi\ra, |\varphi\ra\}\subseteq \IC^n$ satisfying 
(\ref{MP3equal}) for given Hermitian $A, B\in \bM_n$ as follows.

\rm

{\bf Case 1.}
For $n = 2$,  choose $\mu\in \{i, -i\}$
such that $\mu(AB-BA)$ has non-negative $(1,1)$ entry. 
Then the equality holds if $|\psi\ra = |0\ra, |\varphi\ra = |1\ra$.

\smallskip
{\bf Case 2.}
For $n > 2$, determine $\mu \in \{i,-i\}$ such that
$\mu(AB-BA)$  has non-negative $(1,1)$ entry.
Then find a unitary $V = [1] \oplus V_1 \in \bM_n$ such that
the first row of $V^\dag(A-\mu B)V$ only has nonzero entries in the 
$(1,1)$ and $(1,2)$  entries. Then
(\ref{MP3equal}) holds if $|\psi\ra, |\varphi\ra$ to be the first two columns of $V$.

\smallskip
Similarly, for the equality case of 
(\ref{MP6}), suppose $A = (a_{ij}), B = (b_{ij}) \in \bM_n$ are Hermitian.
We can determine $\mu \in \{i,-i\}$ so that the $(1,1)$ entry of 
$\mu(AB-BA)$ is non-negative. Let $W = [1] \oplus W_1$
be a unitary matrix such that
the first column of $W_1$ equals a multiple 
$\frac{|u\ra}{\| |u\ra\|} - \mu \frac{|v\ra}{\|v\ra\|}$ with
$$|u\ra = (a_{21}, \dots, a_{n1})^t \ \hbox{ 
and } \ |v\ra = (b_{21}, \dots, b_{n1})^t.$$
 Let $|\psi\ra, |\varphi\ra$
be the first two columns of $W$. Then 
(\ref{MP6equal}) holds.

\section{Conditions for $\Delta(A)\Delta(B)$ or $\Delta(A)^2 + \Delta(B)^2$ 
attaining 0}

Note that for a pure state $|\psi \ra$,  $\Delta(A) \Delta(B) = 0$ if and only if 
$\Delta(A) = 0$ or $\Delta(B) = 0$, i.e., $A |\psi \ra= \alpha|\psi\ra$ 
or $B|\psi\ra = \beta |\psi\ra$.   For a mixed state  state $\rho$,
$\Delta(A)\Delta(B) =0$ if and only
if $\tr(\tilde  A^2 \rho) = 0$
or $\tr(\tilde B^2 \rho) = 0$, 
i.e., the range spaces of  $\tilde  A^2$ and $\rho$ 
have disjoint support, or the ranges of $\tilde B^2$ and $\rho$
have disjoint support.
Equivalently,   if $U$ is unitary such that
$U^\dag \rho U = D_1 \oplus 0_{n-k}$ for an invertible $D_1\in\bM_k$ 
then   $U^\dag \tilde A^2 U = 0_k \oplus Q_1$ 
or $U^\dag \tilde B^2 U = 0_k \oplus Q_2$. 
If the former case holds, and if $U^\dag \tilde  A U 
= \begin{pmatrix} A_{11} & A_{12} \cr A_{21} & A_{22}\cr\end{pmatrix}$, 
then $\tr(A_{11}^\dag A_{11}) + \tr(A_{21}^\dag A_{12}) = 0$.
So, $\tilde A = 0_k \oplus \tilde Q_1$, i.e., $U^\dag AU = \alpha I_k \oplus A_2$.
A similar analysis also applies if 
$\tr(\tilde B^2\rho) = 0$. Thus, we have the following.

\begin{theorem} \label{4.1} Let  $A, B\in \bM_n$ be 
Hermitian matrices and $\rho\in \bM_n$
be a density matrix.

\smallskip\noindent
{\rm  (a)} We have
$\Delta(A)\Delta(B) = 0$ if and only if $\tilde A \rho = 0$ or $\tilde B \rho = 0$.

\smallskip\noindent
{\rm (b)} We have
$\Delta(A)^2+\Delta(B)^2 = 0$ 
if and only if $\tilde A \rho = \tilde B \rho = 0$.
\end{theorem}

Evidently, if the conditions in (b) hold, then so are the conditions in (a).
Nevertheless, 
if $A = \alpha I_k \oplus A_1$ and $B = \beta I_k \oplus   B_1$, then 
any $\rho = D \oplus 0$ will satisfy (b) and also (a).

\begin{corollary} Let $A, B \in \bM_2$ be Hermitian, and $\rho\in \bM_2$ be a density 
matrix. If $\tilde A\rho = \tilde B \rho = 0$, then 
$A, B$ commute.
As a result, if $A, B$ do not commute, 
then $\Delta(A)^2 + \Delta(B)^2 > 0$. \end{corollary}

\it Proof. \rm  Suppose $\tilde A\rho = \tilde B \rho = 0$.
If $\rho$ is rank 2, then $\tilde A= \tilde B$, while if $\rho$ is rank 1, then $U^\dag \rho U = \diag(1,0)$
for some unitary $U$, and
 $U^\dag (A-\alpha I)U = \diag(0,a), U^\dag(B - \beta I) = \diag(0, b)$. 
 In both cases, $A, B$ commute.
\qed

\section{Conclusion and Discussion}

Matrix techniques were used to give short proofs for lower bounds of $\Delta(A)\Delta(B)$ 
and $\Delta(A)^2 + \Delta(B)^2$ due to Kennard, Robertson, 
Schr\"odinger, Maccone and Pati. 
The saturation conditions were determined.
It would be interesting to extend the techniques to obtain results
for other functions 
$f(\Delta(A), \Delta(B))$.

Our discussions focused on matrices, but
the results and proofs can be easily extended to operators 
acting on general Hilbert spaces. Let us consider the position operator $Q$ and the momentum operator $P$ and their uncertainty relations 
$$
\Delta Q \Delta P \geq \frac{1}{2} |\la \psi|[P, Q] |\psi \ra|
$$
for example. Theorem \ref{2.1} claims that the inequality is saturated when there exists $\vartheta$ such that
$$
(\cos \vartheta Q + i \sin \vartheta P|\psi \ra = (\alpha \cos \vartheta + i \beta \sin \vartheta)|\psi \ra.
$$
The right hand side may be put zero since $\alpha \cos \vartheta$ may be eliminated by shifting the origin of the coordinate while $\beta \sin \vartheta$ may be eliminated by making a Galilei transformation to a moving frame. By acting the eigenvector $\la x|$ of $Q$ on the left hand side, we obtain
$$
\Big(\frac{d}{dx} + c  x \Big) \psi(x) =0,
$$   
where $\psi(x) = \la x|\psi \ra$ and $c= \cot \vartheta$. This equation has a normalized solution 
$$
\psi(x) = \left( \frac{c}{\pi}\right)^{1/4} e^{-c x^2/2}
$$
for any non-vanishing constant $c$. This state saturates the inequality as
$$
 \Delta Q=\sqrt{\frac{1}{2c}},
\ \Delta P=\sqrt{\frac{c}{2}},
\ \frac{1}{2} |\la \psi|[P,Q]|\psi \ra = \frac{1}{2}.
$$

\bigskip\noindent
{\bf Acknowledgment}

Li is an affiliate member of the Institute for Quantum Computing, University of Waterloo.
His research was partially supported by the Simons grant 851334.
Nakahara was 
supported by JSPS Grants-in-Aid for Scientific Research (Grant Number 20K03795). 

Nakahara's affiliation with IQM Quantum Computers is provided for identification purposes only and it is not intended to convey or imply IQM's concurrence with, or support for, the positions, opinions, or viewpoints expressed by the author.

\end{document}